\documentstyle[preprint,aps]{revtex}

\newcommand{\be}{\begin{equation}}
\newcommand{\ee}{\end{equation}}
\newcommand{\bea}{\begin{eqnarray}}
\newcommand{\eea}{\end{eqnarray}}
\def\cela{Ce$_x$La$_{1-x}$M$_3$}
\def\sn{CeSn$_3$}
\def\ind{CeIn$_3$}
\def\pb{CePb$_3$}
\def\pd{CePd$_3$}
\def\b6{CeB$_6$}
\def\ep{\epsilon}
\def\ks{{{\bf k}\sigma}}
\def\jd{|\frac52;\Gamma_7\rangle}
\def\jq{|\frac52;\Gamma_8\rangle}

\begin{document}
\draft


\title{ {\bf\it Ab Initio} Calculation of Crystalline Electric Fields and
Kondo Temperatures in Ce-Compounds
}
\author{
J. E. Han$^{a,b}$,
M. Alouani$^a$, and
D. L. Cox$^{a,b}$
}
\address{
$^a$-Department of Physics, The Ohio State
University, Columbus, Ohio  43210\\
$^b$-Institute for Theoretical Physics, University of
California, Santa Barbara  93106-4030\\
}
\maketitle

\widetext
\begin{abstract}
\noindent
We have calculated the band-$f$ hybridizations for \cela\ compounds 
($x=1$ and $x\rightarrow 0$; M=Pb, In, Sn, Pd)
within the local density approximation and fed this into a non-crossing
approximation for the Anderson impurity model applied to both dilute
and concentrated limits.
Our calculations produce crystalline electric field splittings and Kondo 
temperatures with trends in good agreement with experiment 
and demonstrate the need
for detailed electronic structure 
information on hybridization to describe the diverse
behaviors of these Ce compounds.
\end{abstract}


\narrowtext

A pressing issue in the understanding of strongly interacting
electronic materials is how to produce realistic theoretical
descriptions which encompass both crystalline environment and
symmetry effects, well treated by {\it ab initio} 
electronic structure theory,
with dynamical effects best treated within many body 
formalisms.  A case in point is heavy fermion materials 
with strongly interacting $f$-electron states that give rise to
huge electronic mass enhancements. 
Some understanding of these systems has been reached in Anderson 
model approaches which assume a nearly atomic limit picture for 
$f$-states that hybridize with extended conduction states through 
matrix elements determined from electronic structure (local density
approximation or LDA) 
calculations\cite{gs,nca,cooper,levy,steiner}. 
In particular, Gunnarsson and Sch\"{o}hammer\cite{gs},
have calculated high energy spectra
for a number of cerium based metals 
with a $T=0$ variational method, which however ignored crystal
field effects.  The LDA has been used to estimate hybridization
induced crystalline electric field 
(CEF) splittings in cerium systems\cite{cooper}
without including the strong correlation effects which give
rise to the Kondo effect screening of the $f$-electron magnetic 
moments by itinerant electrons.  Second order perturbation
theory in the direct Coulomb interaction strength $U$ has been employed, and
this gives good estimates for electron mass enhancements  
while only partially capturing Kondo effect physics\cite{steiner}.  

In this work, we present first results for a method which
combines a nonperturbative, finite temperature 
diagrammatic approach for the
Anderson model (the Non-Crossing Approximation) with input
parameters from the LDA.  The NCA can properly generate
hybridization induced CEF splittings while giving an excellent 
description of the Kondo effect.  We report calculations of 
the CEF splittings and Kondo scales $T_K$
in CeM$_3$ (M=Pb, In, Sn, Pd) compounds in which experimental 
$T_K$ values vary with M by nearly three orders of magnitude. 
We have computed energy dependent hybridization matrix elements 
between Ce $f$-states and other conduction states within the
LDA in two limits:\\
1) For the dilute alloy system Ce$_x$La$_{1-x}$M$_3$ 
with $x\rightarrow0$;\\
2) For concentrated CeM$_3$ compounds.  \\
We explain and compare the
methods and results of those approaches. 
We correctly find a stable $\Gamma_7$ doublet CEF 
ground state with small $T_K$ values for \pb\ and \ind\ and 
large $T_K$ values with negligible CEF effects for \sn\ and \pd.  

In our work, the CEF splittings are induced by band-$f$ 
hybridization, which is anticipated to be the dominant contribution\cite{levy}. 
This splitting arises as follows:
a CEF state in Ce-$f^1$ configuration is shifted downward by level 
repulsion through virtual $f^1\to f^0\to f^1$ and $f^1\to f^2\to f^1$ 
charge fluctuations. In the presence of  crystalline anisotropy, different
irreducible representations (irreps) of the point group in the $f^1$ manifold 
receive different shifts. Using this idea in second order perturbation theory, 
Wills and Cooper\cite{cooper}
estimated the contribution of the hybridization-induced CEF splittings
on top of extrapolated point charge contributions\cite{pcm}.
Although the electrostatic potential from the cubic
environment can induce the CEF ({\it i.e. }, in the 
point charge model\cite{pcm}), it is difficult to 
produce a good estimate of this contribution in a metal due 
to conduction electron charge screening\cite{pcm} and metallic covalency. 
We shall neglect point charge contributions in this letter. 

We describe the \cela\ systems in terms of effective impurity Anderson 
models\cite{cepb3,cesn3} in the dilute ($x\to 0$) and concentrated
($x=1$) limits at a site of cubic symmetry relevant to the Cu$_3$Au 
structure. In this Letter, we ignore intersite interaction effects 
such as the anti-ferromagnetism found in 
\ind\cite{cein3} and \pb\cite{cepb3}.
The impurity Anderson Hamiltonian of interest reads
\bea
  H & = & \sum_{\ks}\ep_\ks c^\dagger_\ks c_\ks +\frac{1}{\sqrt{N_s}}
      \sum_{\ks,m}(V_{\ks m}c_\ks f_m^\dagger+h.c.) \nonumber \\
    & + & \sum_m \ep_{fm}f_m^\dagger f_m +
      U\sum_{m<m'}n_{fm}n_{fm'},
\eea 
where $m$ is the label of cubic irrep states, 
{\it e.g.} $m=|J\!=\!\frac52;\Gamma_{7+}\rangle$, $V_{\ks m}$ the hopping matrix
element between conduction electron ($c_\ks$) and $f$-orbital ($f_m$),
 $N_s$ the number of sites, $\ep_{fm}$
the $f$-level energy measured from the Fermi level, $U$ the on-site Coulomb 
repulsion between two $f$-electrons. 

We have used $\ep_{fm}=-2.0$ eV for Hund's ground multiplet ($J$=5/2)
in all the calculations, consistent with
experimental\cite{XPS} and theoretical\cite{ef,why2} values.
The spin-orbit (SO) splitting $\Delta_{\text{SO}}$ 
was read off from the separation between $J=5/2$ and $J=7/2$ 
peaks in the Ce 4$f$ projected densities of states (DOS).  
We find $\Delta_{\rm{SO}}=0.35$ eV for all M, in agreement with
atomic values. 
We set the onsite Coulomb repulsion $U\to \infty$ in our many body 
approximation, though we
partially correct for this as we shall describe below.
In the cubic point group symmetry of the CeM$_3$ compounds, the $J={5\over 2}$
multiplet decomposes into a $\Gamma_7$ magnetic doublet ($\jd$) and $\Gamma_8$
quartet ($\jq$), split by an energy $\Delta_{78}$. 
The $J={7\over 2}$ multiplet splits into $\Gamma_6$ and 
$\Gamma_7$ doublets, and a $\Gamma_8$ quartet. Experimentally, the $\Gamma_7$
doublet lies lowest for M=Pb, In ($\Delta_{78}>0$), 
while no CEF splitting is resolved for M=Sn, Pd.

The hybridization matrix elements are calculated from the LDA 
using the Linearized Muffin-Tin
orbital (LMTO) method in the Atomic Sphere Approximation (ASA) including the so
called combined correction term\cite{lmto}.  We assumed the same
Wigner-Seitz radii for Ce and M (=Pb, In, Sn) and used experimental 
lattice constants.
For \pd, the Wigner-Seitz radius of Ce was set to be 10 \% larger 
than that of Pd\cite{ws}.
We used 165 {\bf k}-points in the irreducible Brillouin zone for the
self-consistent solution with DOS integrations carried out using the tetrahedron
method\cite{tetra}. We set the orbital basis as $s,p,d,f$ for Ce and $s,p,d$
for M-ligands.

In the concentrated limit ($x\rightarrow 1$, {\it i.e.,} CeM$_3$), we define
the hybridization $\Gamma^{\text{med}}_{mm'}(\ep)$, in terms of an
effective {\it impurity} Anderson model, with the hybridization
derived from the overlap between $f$-orbital at 
the origin and the rest of the lattice. (See FIG. 1 (a) and FIG. 2 (a)) 
More specifically 
\be
  \Gamma^{\text{med}}_{mm'}(\ep) = -\text{Im}\sum_{\bf R,R'}
     V_{{\bf R}m}V^*_{{\bf R'}m'}G'({\bf R},{\bf R'},\ep+i\eta),
\ee
where $V_{{\bf R}m}$ is the hopping matrix element between $m$-th $f$-orbital 
and ligand orbital at {\bf R} and $G'({\bf R},{\bf R'},\ep+i\eta)$ the 
Greens' function of ligand electron created at {\bf R} and recovered at 
${\bf R'}$, with the central $f$-orbital excluded, as shown in FIG. 1 (a).
Now the array of ligand- and origin excluded 
$f$-orbitals (with origin excluded) serves 
as an effective {\it static} medium coupled to the
$f$-orbital at the origin. 
Our method follows Gunnarsson {\it et. al.}'s suggestion\cite{gunn}
to interpret the $f$-projected DOS  
as the spectral function of an effective {\it non-interacting} 
resonant level model.  This corresponds to the first iteration of a 
``dynamical mean field theory'' or ``local approximation'' to the
interacting problem\cite{infdrev}, which becomes exact in infinite
spatial dimensions.
We 
obtain the hybridization through Hilbert transformation
as follows:
\bea
  & & \Gamma^{\text{med}}_{mm'}(\ep) = -\frac i2\left( {\bf G}^{LDA}
   (\ep+i\eta)^{-1}-{\bf G}^{LDA}(\ep-i\eta)^{-1}\right)_{mm'} \nonumber\\
  & & G^{LDA}_{mm'}(\ep\pm i\eta) = \int dz
   \frac{\rho^{LDA}_{mm'}(z)}{\ep\pm i\eta-z},
\eea
where $G^{LDA}_{mm'}(\ep\pm i\eta)$ is the 14$\times$14 matrix Green's
function, $\rho^{LDA}_{mm'}(\ep)$ the $f$-PDOS derived from the band
calculation with the LDA. We display the 
calculated $x\to 1$ limit model hybridization in FIG. 2 (a).

In the impurity limit ($x\rightarrow 0$), we calculate the hybridization
between $f$ and {\it bare} ligands, that is,
\be
  \Gamma^{\text{imp}}_{mm'}(\ep) = \pi\sum_\ks V^*_{\ks m}
     V_{\ks m'}\delta(\ep-\ep_\ks ). \label{sum}
\ee
While, in the concentrated limit, the $f$-electron hops
into already-formed bonding states of $f$(origin excluded)-ligand, 
this impurity limit hybridization accounts for the overlap of $f$ and 
pre-bonding ligand states. (See FIG. 1 (b) and FIG. 2 (b)) 
Although this hybridization is calculated for a {\it lattice} Anderson 
Hamiltonian, we can still apply it to impurity limit of \cela\ since the 
Ce-Ce hopping is negligibly small\cite{cece}. The procedure
for computing $V_{\ks m}$ is to set up a resonant level {\it lattice} model
Hamiltonian matrix from the highly non-orthogonal basis of linear muffin-tin
orbitals. We reconstruct the Hamiltonian from eigenvectors and eigenvalues
of the LMTO equation in the {\it one-center} expression\cite{lmto}. Hence, in 
the one-center basis, 
the angular momentum states become trivially orthogonal. After diagonalizing
the ligand sector of the matrix and applying a suitable 
unitary transformation 
to the new ligand basis, 
we can directly read off hybridization matrix elements, $V_{\ks m}$. 
To compute $\Gamma^{\text{imp}}_{mm'}(\ep)$, the Brillouin Zone 
sum of Eq. (\ref{sum})
is calculated with the tetrahedra method\cite{tetra}. In this 
limit, we used lattice constants of LaM$_3$ compounds and readjusted the
Fermi energy such that the ligand bands are filled up with $N_{\text{total}}-1$ 
electrons (which accounts for the lattice-wide removal of the single Ce 
4$f^1$ electron).

As seen in FIG. 2, the $x\to 0$ and $x\to 1$ limit 
hybridizations are almost identical
in the high energy region ($|\epsilon-E_F|\ge 0.5$ eV). 
In the low energy region (inset), the peak 
at $-0.1$ eV in the impurity limit ( FIG. 2 (a)) is pushed down to $-0.2$ 
eV in the effective medium hybridization due to the bonding of Ce-$f$ and 
Pb-$p$ orbitals. Since the Kondo temperatures ($T_K$) depend upon the 
hybridization weight below the Fermi energy ($E_F$), this bonding effect 
can lead to a completely different scale of $T_K$ as $x$ changes.  
For \pb\cite{cepb3} and \sn\cite{cesn3} experimental $T_K$ values are constant 
with $x$, and our calculations (Table 1) show this within reasonable accuracy,
given the exponential sensitivity of $T_K$ to model parameters\cite{medhyb}. 
Another difference between the $x\to 0$ and $x\to 1$ limits 
is the extra structure appearing in the 
$x \to 1$ calculation above $E_F$ (dashed line of FIG. 1(a)) 
which is due to flat $f$-bands. Although this feature above $E_F$ 
is qualitatively different from the $x\to 0$ limit, it contributes little 
to $\Delta_{78}$.

We solved the $U\to \infty$ Anderson model by using the well known
Non-Crossing Approximation (NCA) which gives 
a good quantitative description of  Ce compounds except
for $T\le T_p\ll T_K$ (where $T_p$ is a ``pathology scale'' signalling
breakdown of the approximation)\cite{nca}. To the first 
order expansion in $1/N_g$, with $N_g$ the ground multiplet degeneracy, 
the spectral functions of the $f^0$ and $f^1$ states 
are solved for from coupled self consistent non-linear integral
equations for the $f^0,f^1$ self energies. 
To partially correct for our $U\rightarrow\infty$ approximation, 
we have estimated the contribution 
to $\Delta_{78}$ arising from virtual $f^2$ occupancy 
between $\jd$ and
$\jq$ by employing second order perturbation theory with
$U=6$ eV.  We then added the resulting shifts to the {\it bare}
$f$-level positions in the NCA.  
$\Delta_{78}$ values were inferred 
as the peak position separations between $\jd$ and
$\jq$ $f^1$  spectral functions, which include contributions to all 
orders in $V^2$, as per Levy and Zhang\cite{levy} and in 
contrast to Wills and Cooper\cite{cooper}.  The Kondo temperatures $T_K$
were interpreted as the low temperature splitting between 
the $f^0$ spectral peak and the peak of the lowest $f^1$ CEF state
spectrum.  
$T_K$ roughly depends upon the hybridization
strengths as\cite{tk}
\be
  T_K\approx D_{eff}
    \left( \frac{\Gamma_g}{\pi |\ep_f|}\right) ^{\frac{1}{N_g}}
    \left( \frac{D_{eff}}{\Delta_{78}}\right) ^\frac{N_{ex}}{N_g}
    \exp\left(\frac{\pi\ep_f}{N_g\Gamma_g}\right), \label{tkform}
\ee
where $D_{eff}$ is the effective band width,
$\Gamma_g$ hybridization strength at Fermi energy for
ground multiplet, and $N_g(N_{ex})$ the 
degeneracy of ground(excited) multiplets.

The CeM$_3$ compounds (M=Pb, In, Sn) have the doublet ($\jd$) states as 
the lowest lying multiplet with $\Delta_{78}$ values 
in good agreement with the 
experiments\cite{cein3,cepb3,cesn3}, as listed in Table 1. 
For the heavy fermion systems \ind\ and \pb, 
$\Delta_{78}>0$ ($\Gamma_7$ is stable) and $\Delta_{78}\gg T_K$, both 
in agreement with experiment.
The CEF splitting comes from the larger $\Gamma_7$-hybridization in the 
energy region from 1.0 to 3.0 eV, which results in the doublet ($\Gamma_7$)
ground state for M=In,Pb.
Naively, since $\Delta_{78}\sim V^2$, one would expect the larger
crystal field splitting for \ind\ to correspond to a much larger $T_K$ 
value than for \pb\, which is not seen experimentally.  This common 
reasoning
assumes, however, energy independent hybridization, clearly
not the case here as shown in FIG. 2. 
In detail, the hybridization strength averaged over energy 
of \ind\ exceeds that of \pb\ which  
dictates $\Delta_{78}|_{In}>\Delta_{78}|_{Pb}$, while the smaller \ind\
hybridization near $E_F$ yields a smaller $T_K$ value.
Apparently, detailed hybridization calculations are 
critical for {\it quantititative} 
understanding of real heavy fermion materials. 

The {\it intermediate valence} materials, \sn\ and \pd, have
$\Delta_{78}\ll T_K$ due to their large hybridization, as shown in the table. 
This agrees with experiment which fails to resolve CEF peaks. 
For \pd, the LDA yields an anomalously huge hybridization 
(up to 1.5 eV) between 
Pd-$d$ and Ce-$f$ orbitals below the Fermi energy,
and as a result, the estimated $T_K$ was nearly an order of magnitude 
larger than the experimental value\cite{uofpd}.

The magnetic susceptibility $\chi(T)$ provides a measure of the degree
of screening of the local moments by conduction electrons, for which 
we show our in FIG. 3(a). 
\ind\ and \pb\ showed CEF effects at about 220 and 50 K, 
respectively, without sizable moment screening until the lowest accessible
temperature. Starting with well-localized moment at high temperatures,
$\chi(T)$ deviated from Curie-Weiss behavior; 
$\chi(T)$ of \ind\ crossed over from $|\frac52;
\Gamma_7+\Gamma_8\rangle$ to $\jd$ magnetic moment regime
regaining inverse-$T$ behavior at temperatures (10$\sim$100 K)
well below the CEF splitting. For \pb\ and \ind, the Kondo resonance 
peak begins to emerge (not shown here) near few K 
where the numerics of this calculation becomes unstable. 

For \ind, Lawrence {\it et al.}\cite{cein3} interpreted the 
deviation of $\chi(T)$ from the Curie-Weiss behavior
in the low temperature region as arising from spin-fluctuations with the
characteristic temperature $T_{\rm sf}\sim 50$ K. We believe that this effect
is clearly from the CEF splitting ($\Delta_{78}$=125 K). 
The calculated $T_K$ (few K) is much 
lower than CEF splitting and is inaccessible experimentally due to 
the onset of anti-ferromagnetism at 10 K.
This CEF effect has been checked by comparing the experimental results with the 
Curie-Weiss susceptibility of $\Gamma_7$- and $\Gamma_8-$magnetic moments 
as well as the van Vleck susceptibilities. 
In particular, the experimental low temperature $\chi(T)$ curve shows
saturates only  below 10 K ($\ll T_{\rm sf}$) and
above this is well-fitted to the Curie-Weiss form 
with a characteristic temperature 18 K, which can be
interpreted as a combination of N\'{e}el temperature plus the CEF effect
from the excited quartet field $\Gamma_8$. 

In FIG. 3(b), susceptibility curves of \sn\ are plotted 
for several values of $\ep_f$ with a comparison to experiment\cite{cesn3}. 
Since $\chi(T\to 0)(\sim 1/T_K)$ 
is exponentially sensitive to
$\ep_f$ (see Eq. (4)), we performed 
calculations for several $\ep_f$'s ($-2.0$, $-2.1$, $-2.2$ eV). 
Clearly our calculations bracket but do not fit the experimental data.
since a calculation for $\Delta_{78}$=0 fits the data
well\cite{ekim}, we suspect the source of the disagreement may
be an overestimate of crystal field splitting in our calculation,
placing the effective degeneracy between two ($T_K\ll \Delta_{78}$) and six 
($\Delta_{78}=0$).

In conclusion, we reproduced both the band and many body features
of the Ce-compounds by inputting 
LDA(LMTO-ASA) calculated hybridizations into 
$U\to \infty$ NCA calculations for appropriately defined impurity
Anderson models.  We reproduce well experimental trends in $T_K$,
$\Delta_{78}$, and $\chi(T)$ for the CeM$_3$ series. 
This work provides a starting point for quantitative calculations of realistic
heavy fermion systems at finite temperatures.
Improvement may come through a proper inclusion of $f^2$ dynamics ($U\ne
\infty$), a reliable theory for electrostatic CEF contributions, 
and through self-consistent closure of the many body calculations. 

We are grateful to J. W. Allen, M. Steiner and J. W. Wilkins for 
helpful discussions. This work was supported
by the United States Department of Energy, Office of Basic Energy Sciences, 
Division of Materials Research, and J.H. and D.L.C.
acknowledge the support of NSF grant PHY94-07194 at the Institute for
Theoretical Physics. Supercomputer time was provided by the Ohio
Supercomputer Center.

\begin{figure}

\caption{Schematic diagrams of hopping of $f$-electrons to band states
in \cela\ systems.
(a) In the $x=1$ limit, the center $f$-orbital couples the rest of the
lattice (enclosed in dashed line), {\it i.e.,} effective medium. 
The $f$-orbital is connected to 
ligand orbitals ($c_i$) with $V_{im}$ ($i$ lattice site, $m$ the CEF index).
The lattice of $f$ orbitals (origin excluded) are treated as producing a
{\it static} mean field coupling to ligands in the LDA calculation.
(b) In the impurity limit ($x\rightarrow 0$), the $f$-orbital (at center)
couples to the {\it bare} ligand bands. 
}
\label{fig1}
\vfil

\caption{Hybridizations of \cela\ calculated in two different
limits of $x$.
(a) $x=1$ limit.
Although $\Gamma_7(J=5/2)$ (thick line) and $\Gamma_8(J=5/2)$ (thin
line)-hybridizations are comparable, large hybridization 
($f^1\leftrightarrow f^0$ fluctuation) for $\Gamma_7$ 
near 1.0-3.0 eV pushes the doublet below the quartet. The spikes appearing
above the Fermi energy (dashed line) are due to the coupling of $f$-orbital 
at origin to the bonding (anti-bonding) states of ligand and 
neighboring $f$-orbitals.
(b) $x\to 0$ limit. 
Curves are almost identical to (a) except
near the Fermi energy. Note that the peak at $-0.1$ eV (inset) is pushed
down to $-0.2$ eV (inset (a)) due to $f$-ligand coupling. 
}
\label{fig2}
\vfil

\caption{Calculated magnetic susceptibility 
$\chi(T)$-vs.-$T$ for CeM$_3$ compounds. 
(a) M dependence of $\chi(T)$; see text for discussion.  
(b) Magnetic susceptibilities of \sn\ for different $f$-level 
positions, $\ep_f$. }
\label{fig3}

\end{figure}

\begin{table}
\caption{Table of the electronic configuration of ligand atom M (E.C.)
ground CEF multiplet $\Gamma_{grd}$, CEF splittings 
$\Delta_{78}$ and Kondo temperatures $T_K$ for \cela\ systems.
Units are in Kelvin. Positive $\Delta_{78}$ indicates a stable 
$\Gamma_7$ ground doublet on the Ce sites. 
(*: For mixed valence systems, CEF's are not experimentally 
deducible since $\Delta_{78}\ll T_K$). 
}
\label{table1}
\begin{tabular}{llcccc}
  M & E.C. & $\Gamma_{\text{grd}}$ & 
  $\Delta^{\text{exp}}_{78}/T^{\text{exp}}_K$ & 
  $\Delta^{x=1}_{78}/T^{x=1}_K$ & 
  $\Delta^{x\rightarrow 0}_{78}/T^{x\rightarrow 0}_K$ \\ \hline
 Pb & $6s^2 6p^2$            & $\Gamma_7$ & 
     67/3   &  50/($\lesssim 1) $ & 40/23 \\
 In & $4d^{10}5s^2 5p^1$    & $\Gamma_7$ & 
     183/($<11$) & 125/($< 1.0$) & 80/1.1 \\
 Sn & $4d^{10}5s^2 5p^2$     & $\Gamma_7^*$ & 
     -/450 & 159/400 & 176/238 \\
 Pd & $4d^{10}$              & $\Gamma_8^*$ & 
     -/700 & 215/3210 & 314/2600 \\
\end{tabular}
\end{table}


\begin{references}
\bibitem{why2} We chose the -2.0 eV value based upon
the interconfiguration energy difference curves
of Ref. \cite{ef} at the Wigner-Seitz cell value $r_{WS}(Ce)$
of elemental Ce.  Our CeM$_3$ LDA $r_{WS}$ values vary little
from $r_{WS}(Ce)$ with M.
\bibitem{ws} We plot each atomic potential as a function of 
	 atomic radius and determine by inspection the ratio of the 
         Wigner-Seitz radii of Ce and M when the potential becomes flat.
\bibitem{cece} We explicitly checked that the hyridization is dominated by
         ligand($s,p$ for Pb, In, Sn or $d$ for Pd)-$f(Ce)$ overlap. 
         The Ce-Ce hopping ($d$-$f$, as well as $f$-$f$) is more than
         an order of magnitude smaller.
\bibitem{medhyb} The impurity limit calculation (FIG. 1 (b)) does not include
             the effect of $f$-orbital of La atoms which would reduce $T_K$ 
	     due to the extra downward shift of the peak at $-0.1$ eV 
	     (inset of FIG. 2 (b)).
\bibitem{tk} The (difficult) dynamic
inclusion of finite $U$ will 
modify $T_K$ values due to the virtual exchange
processes mediated by the $f^2$ configuration;  practically,
this will require recalibration of the $\epsilon_f$
values we input into the NCA.
\bibitem{uofpd} The overestimated $\Gamma(\epsilon)$ values for
\pd\ may arise from the LDA underestimation of the $d(Pd)$-band width.
\end{references}
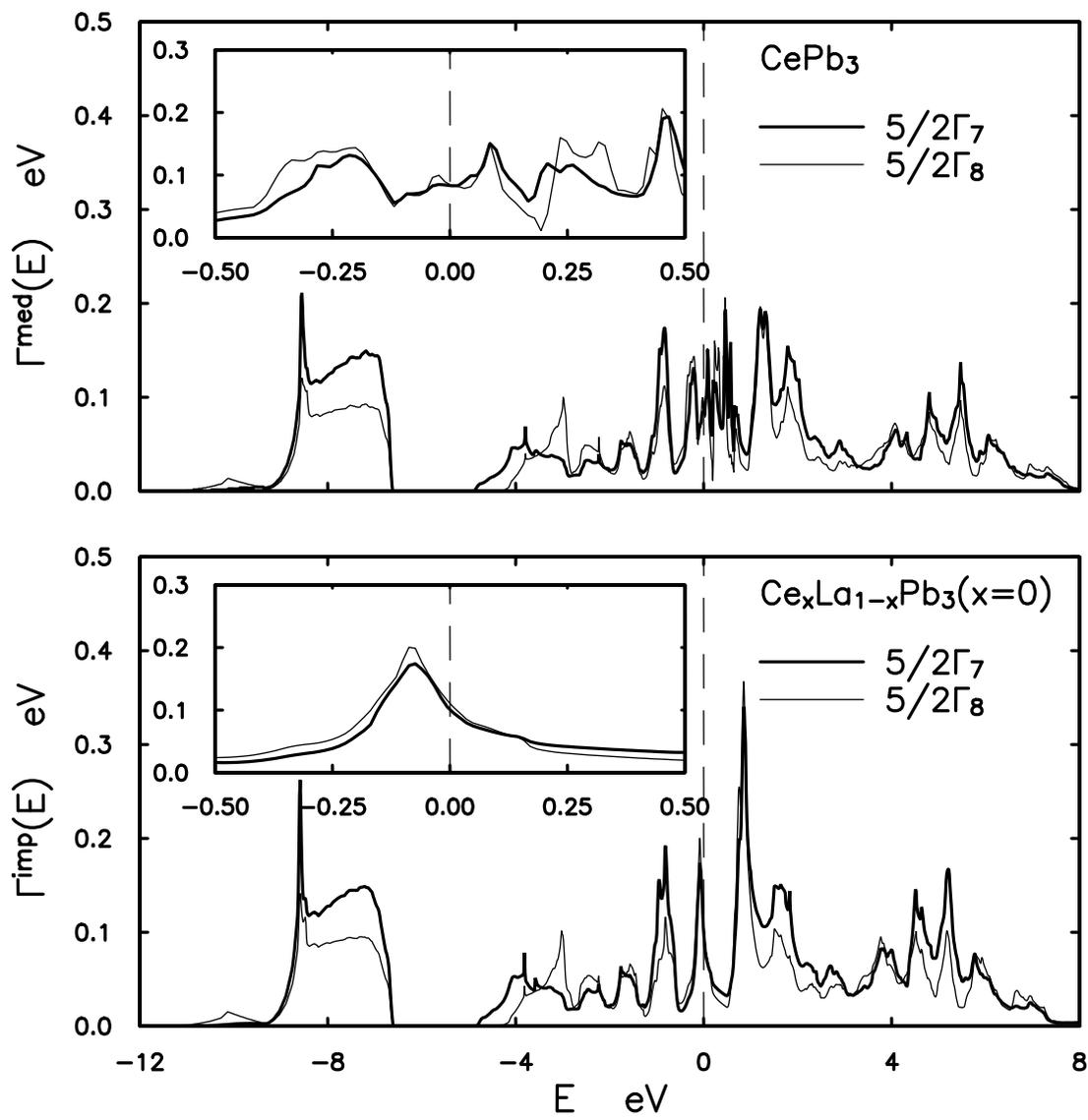
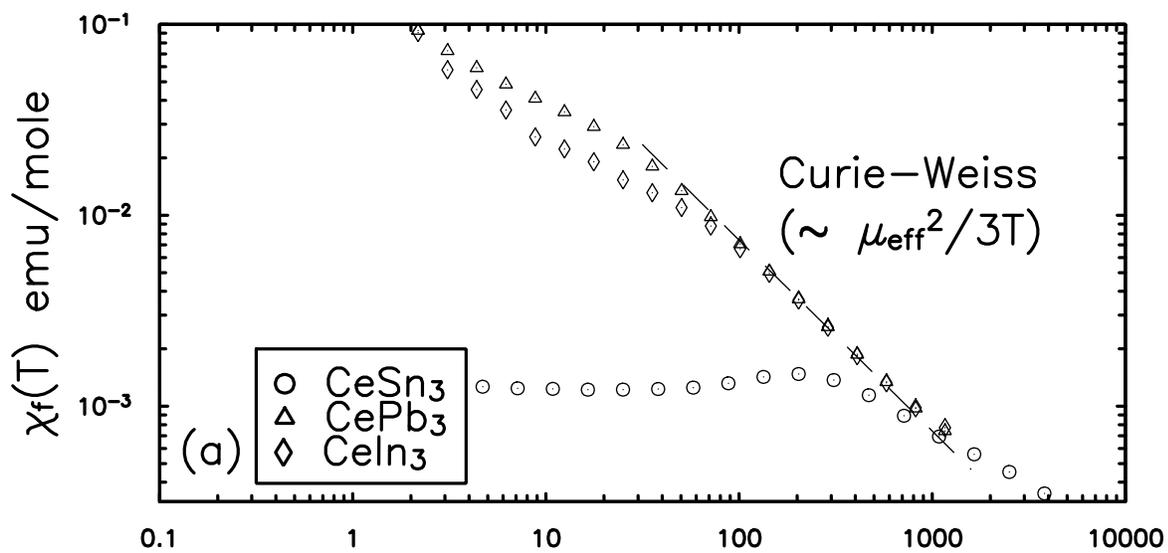
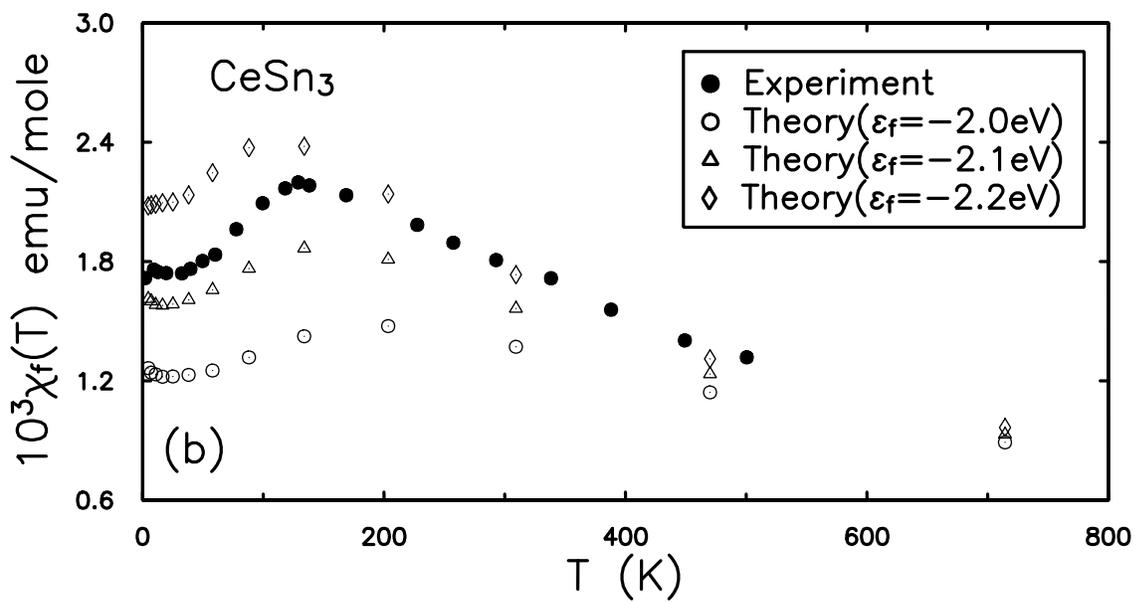
\end{document}